\documentclass{article}

\usepackage[english]{babel}

\usepackage[letterpaper,top=2cm,bottom=2cm,left=3cm,right=3cm,marginparwidth=1.75cm]{geometry}

\usepackage{authblk}
\usepackage{amsmath}
\usepackage{amsfonts}
\usepackage{graphicx}
\usepackage[colorlinks=true, allcolors=blue]{hyperref}

\usepackage{listings}
\usepackage{forest}
\usepackage{tikz-qtree}
\usepackage[linesnumbered,lined,boxed,commentsnumbered]{algorithm2e}
\usepackage{balance}
\usepackage{multirow}
\usepackage{caption}
\usepackage{subcaption}

\usepackage{fancyvrb} 

\DeclareMathDelimiter{(}{\mathopen} {operators}{"28}{largesymbols}{"00}
\DeclareMathDelimiter{)}{\mathclose}{operators}{"29}{largesymbols}{"01}

\newcommand{\true}{\texttt{True}} 
\newcommand{\false}{\texttt{False}}

\newcommand{\ord}{\bigtriangleup}

\newcommand{\minl}{\texttt{MinL}}
\newcommand{\len}{\texttt{Len}}
\newcommand{\first}{\texttt{First}}
\newcommand{\last}{\texttt{Last}}
\newcommand{\node}{\texttt{Node}}
\newcommand{\edge}{\texttt{Edge}}
\newcommand{\prop}{\texttt{Prop}}
\newcommand{\lab}{\texttt{Label}}

\newcommand{\s}{\mathbb{S}}
\newcommand{\nodes}{\texttt{Nodes}}
\newcommand{\edges}{\texttt{Edges}}

\newcommand{\ptog}{\alpha}
\newcommand{\gtop}{\beta}

\newcommand{\code}[1]{{\texttt{#1}}}

\def\checkmark{\tikz\fill[scale=0.4](0,.35) -- (.25,0) -- (1,.7) -- (.25,.15) -- cycle;} 

\newtheorem{definition}{Definition}[section]

\title{Path-based Algebraic Foundations of Graph Query Languages}

\author[1]{Renzo Angles \thanks{Corresponding Author: renzoangles@gmail.com}}
\author[2]{Angela Bonifati}
\author[1]{Roberto García}
\author[3]{Domagoj Vrgo\v{c}}
\affil[1]{Universidad de
Talca  \& IMFD Chile}
\affil[2]{Lyon 1 Univ., Liris CNRS \& IUF}
\affil[3]{PUC Chile \& IMFD Chile}

\begin{document}
\maketitle

\begin{abstract}
Graph databases are gaining momentum thanks to the flexibility and expressiveness of their data models and query languages. A standardization activity driven by the ISO/IEC standardization body is also ongoing and has already conducted to the specification of the first versions of two standard graph query languages, namely SQL/PGQ and GQL, respectively in 2023 and 2024. Apart from the standards, there exists a panoply of concrete graph query languages provided by current graph database systems, each offering different query features. A common limitation of current graph query engines is the absence of an algebraic approach for evaluating path queries. To address this, we introduce an abstract algebra for evaluating path queries, allowing paths to be treated as first-class entities within the query processing pipeline. We demonstrate that our algebra can express a core fragment of path queries defined in GQL and SQL/PGQ, thereby serving as a formal framework for studying both standards and supporting their implementation in current graph database systems. We also show that evaluation trees for path algebra expressions can function as logical plans for evaluating path queries and enable the application of query optimization techniques. Our algebraic framework has the potential to act as a lingua franca for path query evaluation, enabling different implementations to be expressed and compared.
\end{abstract}

\section{Introduction}
Graph databases are becoming a widely spread technology, leveraging the property graph data model, and exhibiting great expressiveness and computational power \cite{Bonifati2018}. The success of graph data systems such as Neo4j, TigerGraph, MemGraph, Oracle PGX, AWS Neptune and RedisGraph had led to a standardization activity around graph query languages, carried out by the ISO/IEC standardization body. The ISO/IEC has already finalized the first version of SQL/PGQ \cite{SQLPGQ2023} as part of the 2023 version of the SQL standard and has recently finalized GQL \cite{GQL2024}, a native graph query language that will eventually not only return tables but also paths and graphs. 


Finding and returning paths is a fundamental part of every graph query language as witnessed by the rich set of features for manipulating paths in the SQL/PGQ and GQL standard. However, while the ISO standards do prescribe mechanisms for path manipulation, current engines are severely lagging in their implementation. Indeed, to the best of our knowledge, there is currently no engine that fully supports path features prescribed by the two standards and most solutions deploy their own semantic of syntactic specification of what is allowed in terms of path queries~\cite{Farias2023ARXIV}. We believe that one reason for this is the lack of a common algebra that allows expressing the multitude of path query features required by graph engines and prescribed by the SQL/PGQ and GQL standards.

Therefore, in this paper we lay the foundations of a path-based algebraic framework for evaluating path queries. Our effort is relevant from both a theoretical viewpoint and a system perspective. Indeed, a standard graph query algebra is missing while being a core component of the next-generation graph ecosystems and their user cases \cite{SakrBVIAAAABBDV21}.
Our framework is \emph{expressive} as it encompasses the fundamental path features of current graph query languages and the ISO standards while precisely formalizing their semantics. It also offers \emph{query composability}, allowing to specify algebraic expressions that can be arbitrarily nested and returns paths that are consumed by other queries. It also exhibits  \emph{strict adherence} to the standard graph query languages in terms of the covered query operators and the different variants of the query semantics it can support. It also embodies a blueprint for the \emph{algebra-based implementation} of graph queries across systems, since it directly compiles into logical plans for executing graph queries, paving the way to the final adoption of the graph query language standards themselves. Additionally, our formalization of the path-based algebra \emph{anticipates} the future versions of the query language standards, expressing several natural properties outside of their current scope. 

To illustrate the expressiveness of our path-based algebraic framework, we introduce an example in the following. 

\begin{figure}[t!]
  \centering
  \includegraphics[width=\linewidth]{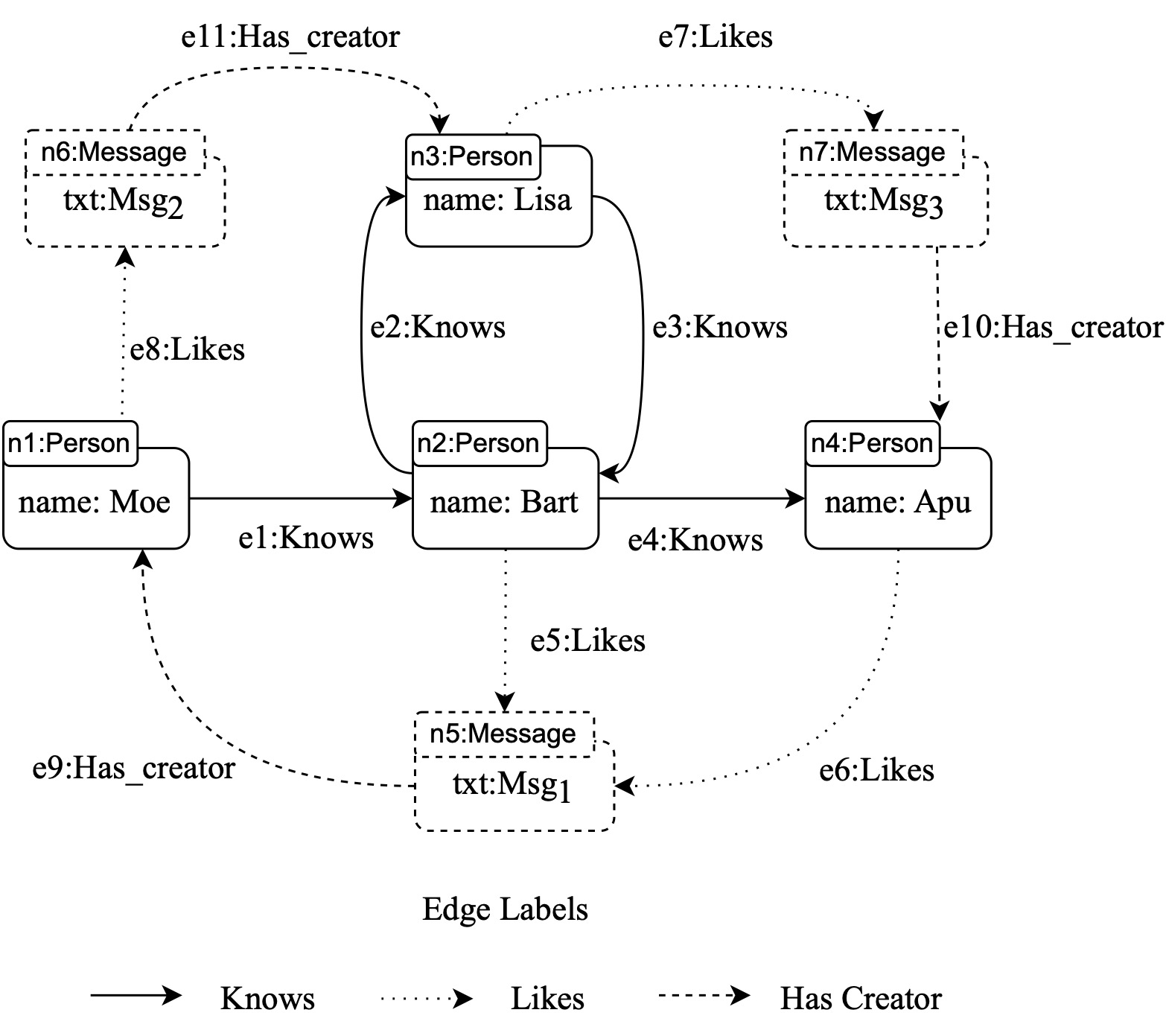}
  \caption{A graph representing a social network (drawn from the LDBC SNB benchmark).}
  \label{fig:exampleGraph}
\end{figure}

\paragraph{Path algebra by example} 
Consider the property graph shown in Figure \ref{fig:exampleGraph}, which is a snippet of the LDBC Social Network Benchmark graph \cite{SzarnyasWSSBWZB22}, a popular benchmark for property graph databases. The graph relates Persons and Messages, connected through relationships Knows, Likes and Has\_Creator. An essential characteristic of this graph is the capability to employ recursion, due to the presence of cycles. One can observe a double cycle in Figure \ref{fig:exampleGraph}, with an inner cycle involving $Knows$ edges and an outer cycle traversing the concatenation of edges labeled as $Likes$ and $Has\_creator$. 

An example of path query (in GQL-like syntax) leveraging the cyclic structure of the underlying graph data is reported below. 
The query computes all the paths from the node $n1$ (e.g. ``Moe'') to the node $n4$ (e.g. ``Apu''), either across the inner cycle with label \code{Knows} or across the outer cycle with labels \code{Likes} and \code{Has\_creator}. 

{ \small
\begin{verbatim}
  MATCH  p = (?x {name:"Moe"})-[(:Knows+)|
             (:Likes/:Has_creator)+]->(?y {name:"Apu"}) 
\end{verbatim}
}

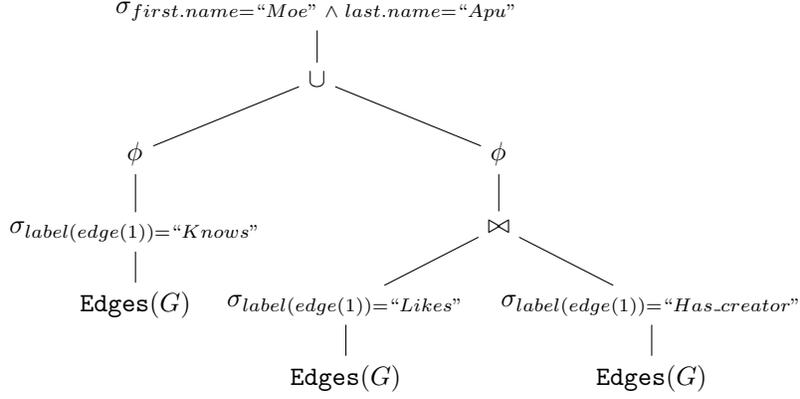
\begin{figure}[t!]
  \centering
  \begin{forest}
    [$\sigma_{first.name = ``Moe" \; \wedge \; last.name = ``Apu"}$
        [$\cup$
            [$\phi_{}$
                [$\sigma_{label(edge(1)) = ``Knows"}$
                    [$\edges(G)$]
                ]
            ]
            [$\phi_{}$
                [$\bowtie$
                    [$\sigma_{label(edge(1)) = ``Likes"}$
                    [$\edges(G)$]
                    ]
                    [$\sigma_{label(edge(1)) = ``Has\_creator"}$
                    [$\edges(G)$]
                    ]
                ] 
            ]
        ]
    ]   
\end{forest}
  \caption{The algebraic plan of a recursive graph query ($\phi_{}$ being the recursive operator corresponding to Kleene plus).}
  \label{fig:qtreeRecAlgebra}
\end{figure}

In this paper, we introduce a comprehensive path algebra that allows expressing such queries in an abstract fashion. Moreover, each evaluation tree of a path algebra expression can be viewed as a logical plan for evaluating queries such as the one above. An example of an evaluation tree for our query is given in Figure \ref{fig:qtreeRecAlgebra}.

In general, our algebra mimics standard relational algebra in terms of symbols used, but it operates on sets of paths instead of relations. As such, it uses the set of nodes and the set of edges (i.e. paths of length zero and one respectively) as its atoms, and combines them to construct and filter out paths. 
Our algebra is divided into three groups: core path algebra, recursive path algebra, and extended path algebra. 

The core algebra includes three operators: selection, join and union.
The \emph{selection} operator ($\sigma$) filters a set of paths according to a specific selection condition (e.g. $label(edge(1))=``Likes"$, to access the label of the first edge in a path, or $first.name=``Moe"$, to access the $name$ attribute of the first node along a path). 
The \emph{union} operator ($\cup$) computes the union of two sets of paths, eliminating duplicates.
The \emph{join} operator ($\bowtie$) combines paths from two sets by generating a new path for each pair of paths that share the same final and initial nodes, thus mimicking the concatenation of paths. 

The recursive path algebra is integrated by the recursive operator ($\phi$), which computes a recursive self-join over a set of paths, allowing the construction of paths of arbitrary length.
Returning to the example shown in Figure \ref{fig:qtreeRecAlgebra}, the evaluation of the query tree over the graph in Figure \ref{fig:exampleGraph} will produce infinite results; it is given by the inner loop formed by the label \code{Knows}, and the outer loop formed by the concatenation of the label \code{Likes} with \code{Has\_Creator}. 
Indeed, the underlying default semantics of the $\phi$ operator is the arbitrary paths semantics (which corresponds to the default WALK semantics in GQL), computing all possible paths between pairs of nodes. 
In this case, due to the presence of the two cycles in the graph in Figure \ref{fig:exampleGraph}, the query will never halt, since it can keep on looping and returning longer and longer paths. 

To cope with the issue of infinite results, GQL and SQL/PGQ impose a tight policy on paths that can be returned through the concept of \emph{restrictors}, that control the type of paths that are matched to the query (for instance SHORTEST WALKS or SIMPLE paths). Our algebra mimics this behavior by specializing the $\phi$ operator in accordance to different semantic restrictions that need to be imposed.
Specifically, in addition to the arbitrary semantics ($\phi_{Walk}$), our algebra provides recursive operators for acyclic ($\phi_{Acyclic}$), simple path ($\phi_{Simple}$), trail path ($\phi_{Trail}$) and shortest path ($\phi_{Shortest}$) semantics. All these semantics are included in the core pattern matching fragment of both GQL and SQL/PGQ, which is common to the two standards. 
Hence, if we change the recursive operators in our example query tree with $\phi_{Simple}$, then the result of the query will only contain the following two paths: 
\begin{align*}
path_1 & = (n_1,e_1,n_2,e_4,n_4) \\
path_2 & = (n_1,e_8,n_6,e_{11},n_3,e_7,n_7,e_{10},n_4),
\end{align*}
where we denote a path as an interchanging sequence of nodes and edges, starting and ending with a node.

The extended algebra introduces the notion of solution space and defines three operators: group-by, order-by and projection. 
The group-by operator generates a solution space from a set of paths, where the paths are organized in partitions and groups.
The order-by operator sort the paths, the groups and the partitions of a solution space.
The projection operator returns a set of paths extracted from a solution space according to given criteria.  
The components of the extended algebra were designed to support different types of \emph{selectors}, which is a novel feature introduced in GQL and SQL/PGQ.

A key feature of our algebra is query composability, as a set of paths serves as the primary data structure for input and output in the algebra operators (with solution spaces as secondary data structures).
It is important to highlight that current graph query languages are unable to manipulate set of paths, and query composability is often lost when returning paths in graph queries. 

Overall, our contributions can be summarized as follows:
\begin{itemize}

\item We introduce an abstract algebra for evaluating regular path queries, allowing paths to be treated as first-class entities within the query processing pipeline.

\item 
We demonstrate that our algebra can express a core fragment of path queries defined in GQL and SQL/PGQ, and can therefore serve as a formal framework for studying both standards.

\item
We provide precise and concrete semantics for the selectors and restrictors introduced in GQL and SQL/PGQ.
Additionally, we include several natural graph operators missing from the two proposals, providing space for future additions to the standard.

\item
We also show that evaluation trees for path algebra expressions can function as logical plans for evaluating path queries and enable the application of query optimization techniques.
Concretely, once we have an algorithm for each operator in the algebra, a sound proof of concept implementation of the GQL and SQL/PGQ standards can be provided with ease. 

\item We provide an open-source parser of the algebra and we make it publicly available for the wider community. 
    
\item Our algebraic framework has the potential to act as a lingua franca for path query evaluation, enabling different implementations to be expressed and compared.

\end{itemize}

\section{Preliminaries}
\label{sec:core}
The path algebra proposed in this article has been designed for returning and manipulating sets of paths. In this sense, each algebra operator takes one or two sets of paths as input, and its evaluation returns a single set of paths.
Next we introduce basic concepts associated to property graphs and paths.

\subsection{Property graphs}
Informally, a property graph is a directed labelled multigraph with the special characteristic that each node or edge could maintain a (possibly empty) set of property-value pairs \cite{91401}. From a data modeling point of view, a node represents an entity, an edge represents a relationship between entities, and a property represents a specific feature of an entity or relationship.

Formally, let \textbf{O} be an infinite set of object identifiers, \textbf{L} be an infinite set of labels, \textbf{P} be an infinite set of properties, and \textbf{V} be an infinite set of values.
A \emph{property graph} can be defined as follows. 

\begin{definition}
\label{def:PG}
A property graph (PG) is a tuple G = $(N,E,\rho,\lambda,\nu )$ where:
\begin{enumerate}
\item $N \subset \textbf{O}$ is a finite set of node identifiers;
\item $E \subset \textbf{O}$ is a finite set of edge identifiers where $N \cap E  = \emptyset$;
\item $\rho : E \to (N \times N)$ is a total function that defines the pairs of nodes connected by each edge; 
\item $\lambda : (N \cup E) \rightharpoonup \textbf{}$ \textbf{L} is a partial function that assigns a single labels to some nodes and edges;
\item $\nu: (N \cup E)\ \times$  \textbf{P} $\rightharpoonup$ \textbf{V} is a partial function that enables the definition of properties for nodes and edges.
\end{enumerate}
\end{definition}

As an illustration, consider again the graph $G = (N,E,\rho,\lambda,\nu)$ from Figure \ref{fig:qtreeRecAlgebra}. 
According to the above definition, we will have that
$N = \{n_1,\dots,n_7\}$ is the set of node identifiers, and $E = \{e_1,\dots,e_{11}\}$ is the set of edge identifiers.
Given the node $n_1$, we have that $\lambda(n_1) = \text{Person}$ is the label of $n_1$.
Given an object $o$ (node or edge), if $o \not \in range(\lambda)$ then $\lambda(o) = \emptyset$.
Given the edge $e_1$, if $\rho(e_1)$ = $(n_1,n_2)$ then $n_1$ and $n_2$ are the source node and the target node of $e$ respectively. 
The function $\nu(obj,prop) = val$ allows us to assign the value $val$ to the property $prop$ of an object $obj$. For example, if $\nu(n_1,\code{name}) = ``Moe"$, then $``Moe"$ is the value of the property $\code{name}$ for the node $n_1$.

\subsection{Paths}
A path $p$ in a property graph $G = (N,E,\rho,\lambda,\nu)$ is a sequence of node and edge identifiers of the form 
\[ p = (n_1, e_1, n_2, e_2, n_3, e_3, \dots, e_{k},n_{k+1} )\]
where
$k \geq 0$, $n_i \in N$, $e_i \in E$, and $\rho(e_i)=(n_{i},n_{i+1})$ for $1 \leq i \leq k+1$.
The last condition ensures that, for each pair of edges $e_i$, $e_j$ in $p$, the target node of $e_i$ is equal to the source node of $e_j$.  
The label of $p$, denoted $\lambda(p)$, correspond to a string formed by the concatenation of the edge labels occurring in $p$, i.e. $\lambda(e_1) \dotsi \lambda(e_{k})$.

The length of a path $p$ is the number of edge identifiers in $p$.
Note that a path of length zero is formed by a single node (without edges).  
Given the graph $G$, the function 
$\nodes(G) = N$ returns the set of nodes (i.e. paths of length zero) and function $\edges(G) = E$ returns the set of edges (i.e. paths of length one).  

Given two paths $p_1$ and $p_2$ are equal if they have the same sequence of node and edge identifiers. A path is called \emph{acyclic}, if $n_i\neq n_j$, for all $i\neq j$ and it is called \emph{simple} if $n_i\neq n_j$ for all $i \neq j$, except that we allow $n_1 = n_{k+1}$, meaning that start and end node can be the same. A path is a \emph{trail}, if $e_i \neq e_j$, for all $i \neq j$. Finally, we remark that, following theoretical graph literature, GQL and SQL/PGQ use the term \emph{walk} to indicate an arbitrary path.

\subsection{SQL/PGQ and GQL}
\label{sec:gql}
In this section, we provide a concise recap of the formalization of GQL \cite{GQL2024} and SQL/PGQ \cite{SQLPGQ2023} path queries. 
Path queries in both standards are based on \emph{path patterns}, which are an extension of regular path queries~\cite{CruzMW87}(RPQs), which are expressions of the form $(x,\texttt{regex},y)$, with $x,y$ being variables or constants, and \texttt{regex} a regular expression. Such a query then returns all pairs of nodes $(n,n')$ in a property graph that are linked by a path whose edge labels form a word matching \texttt{regex}. 


While in the research literature RPQs only look for nodes and not for paths~\cite{AnglesABHRV17}, in GQL and SQL/PGQ, one is also interested to retrieve paths witnessing these connections.
Of course, as illustrated in the introductory example, in the presence of cycles there is a potentially infinite number of such paths. To cope with this issue, GQL and SQL/PGQ introduce \emph{selectors} and \emph{restrictors} as a way to select the paths to be returned, and to specify the semantics used for computing the paths, respectively. 
For example, consider the following path query in GQL,
{ 
\small
\begin{verbatim}
  ANY SHORTEST WALK p = (x)-[:Knows]->+(y),
\end{verbatim}
}
\noindent
where \code{ANY SHORTEST} is the selector clause, and \code{WALK} is the restrictor clause.  
In this case, the restrictor indicates that the query will compute the paths between any pair of people, connected by edges labeled \code{Knows}, one or more times, without any kind of restriction (i.e arbitrary path semantics).
Additionally, the selector indicates that among all the retrieved paths, the query must return just a single shortest path, selected randomly. 
The allowed selectors and restrictors, and their corresponding semantics are presented in Table~\ref{tab:selectors} and Table~\ref{tab:restrictors} respectively.

Following~\cite{Farias2023ARXIV,Francis2023ICDT}, we can define a \emph{path query} in GQL and SQL/PGQ as an expression of the form
$$\textsf{selector}?\  \textsf{restrictor}\  (x,\texttt{regex},y).$$
In general terms, a query will return all the paths that match the specified selector-restrictor combination while at the same time being an answer to the underlying path pattern. We remark that the selector part is optional, with the restriction that for the WALK restrictor the selector must be specified in order to ensure a finite answer set.

\begin{table*}[h!]
\centering
\begin{tabular}{|l|l|}
\hline
\textbf{Expression} & \textbf{Informal semantics} \\ \hline
\code{ALL} 
& Returns all paths, for every group, for every partition. 
\\ \hline

\code{ANY SHORTEST} 
& Returns one path with shortest length from each partition. \\
& \textbf{Non-deterministic}.
\\ \hline
\code{ALL SHORTEST} 
& Returns all paths in each partition  that have the minimal length in the \\
& partition. \textbf{Deterministic}.
\\ \hline

\code{ANY} 
& Returns one path in each partition arbitrarily. \textbf{Non-deterministic}.
\\ \hline

\code{ANY} $k$ 
& Returns arbitrary $k$ paths in each partition (if fewer than $k$,  then all \\
& are retained). \textbf{Non-deterministic}.
\\ \hline

\code{SHORTEST} $k$ 
& Returns the shortest  $k$ paths (if fewer than $k$, then all  are retained). \\
& \textbf{Non- deterministic}.
\\ \hline

\code{SHORTEST} $k$ \code{GROUP} 
& Partitions by endpoints, sorts each partition by path length, groups \\
& paths with the same length, then returns all paths in the first $k$ groups \\
& from each partition (if fewer than $k$, then all are retained). \textbf{Deterministic}. 
\\ \hline
\end{tabular}
\caption{Selectors in GQL}
\label{tab:selectors}
\end{table*}

\begin{table*}[h!]
\centering
\begin{tabular}{|l|l|}
\hline
\textbf{Expression}& \textbf{Informal semantics} 
\\ \hline
\code{WALK} & Is the default option, corresponding to the absence of any filtering.                                      \\ \hline
\code{TRAIL} & Returns paths that do not have any repeated edges.
\\ \hline
\code{ACYCLIC} &  Returns paths that do not have any repeated nodes.
\\ \hline
\code{SIMPLE} 
& Returns paths with no repeated nodes, except for the first and last node if they \\
& are the same.  
\\ \hline
\end{tabular}
\caption{Restrictors in GQL}
\label{tab:restrictors}
\end{table*}

In addition to plain path queries, GQL and SQL/PGQ allow concatenating two path queries into a sequence. For instance we can write $s\ r [s_1\ r_1\ (x,\texttt{regex1},y)] \cdot [s_2\ r_2\ (z,\texttt{regex2},w)]$, where $s,s_1,s_2$ are selectors, $r,r_1,r_2$ restrictors, and the query basically concatenates (when possible) paths in the answer of $s_1\ r_1\ (x,\texttt{regex1},y)$ and $s_2\ r_2\ (z,\texttt{regex2},w)$ and applies the $s\ r$ selector-restrictor combination to that set. This in particular means that we can ask for all trails connecting nodes $n_1$ and $n_2$, then all shortest walks connecting $n_2$ to $n_3$, and require that the entire concatenated path between $n_1$ and $n_3$ be a shortest trail. Another option allowed by GQL is taking an union of such answer sets, with the usual set-union semantics.

Finally, we remark that some functionalities of GQL such as group variables~\cite{Francis2023ICDT} are not covered in this paper, but given that these are used to collect nodes or edges along a path into a list, incorporating them into our framework is rather straightforward.

\section{Core Path Algebra}
\label{sec:core}
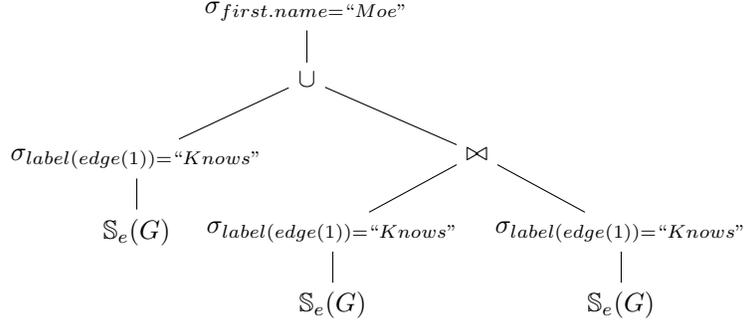
\begin{figure}[t!]
  \centering
  \begin{forest}
    [$\sigma_{first.name = ``Moe"}$
        [$\cup$
            [$\sigma_{label(edge(1)) = ``Knows"}$
                [$\s_e(G)$]
            ]
            [$\bowtie$
                [$\sigma_{label(edge(1)) = ``Knows"}$
                    [$\s_e(G)$]
                ]
                [$\sigma_{label(edge(1)) = ``Knows"}$
                    [$\s_e(G)$]
                ]
            ] 
        ]
    ]
\end{forest}
  \caption{Example of query tree obtained from a core path algebra expression.}
  \label{fig:qtreeBALgebra}
\end{figure}

Given our sample graph (Figure \ref{fig:exampleGraph}), suppose that we would like to obtain the paths containing the friends and the friends-of-friends of \code{"Moe"}, i.e. the 1-hop and 2-hop paths.
This question can be answered by using the following GQL-like query:
{ \small
\begin{verbatim}
  MATCH p = (?x {name:"Moe"})-[Knows|(Knows/Knows)]->(y).
\end{verbatim}
}
In the above expression: 
\code{(x \{name:"Moe"\})} denotes the source node,
\code{[Knows|(Knows/Knows)]} is a regular expression,
\code{(y)} denotes the target node, and
\code{p} is a variable used to contain the resulting paths.
The above declarative query can be transformed into an algebra expression whose evaluation tree is shown in Figure \ref{fig:qtreeBALgebra}. 
Next we explain the operators that conform the core of the path algebra proposed in this article. These are: selection, join  and union.

Given a set of paths $S$, the \emph{selection} operator ($\sigma$) allows filtering the paths in $S$ according to a filter condition. In our example, the algebra expression $\sigma_{label(edge(1)) = ``Knows"}$ filters the paths in $\s_e(G)$ (i.e. the paths of length one in $G$) such that, each path satisfies that its first edge has label \code{``Knows"}. 

Given two sets of paths $\s_1$ and $\s_2$, the \emph{join} operator ($\bowtie$) returns a set of new paths where each new path is the result of concatenating a path $p_1$ from $\s_1$ and a path $p_2$ from $\s_2$ such that the last node of $p_1$ is equal to the first node of $p_2$.
In our example, the join operator for paths is used to obtain the paths having the structure $(n_1,``Knows",n_2,``Knows",n_3)$.

Following the usual semantics in set theory, the \emph{union} operator ($\cup$) combines two sets of paths into a single set of paths that includes all the paths from the input sets. In our example, we use the union operator to combine the paths of the form $(n_1,``Knows", n_2)$ with the paths of the form $(n_1,``Knows",n_2,``Knows", n_3)$.  

Finally, the selection expression $\sigma_{first.name = ``Moe"}$ in the root node of the evaluation tree allows filtering the paths returned by the union operator, such that each final path satisfies that its first node has a property \code{``name''} with value \code{``Moe''}.

Note that the core algebra is closed under set of paths. i.e. the input and the output of every operator is always a set of paths. It is very important because it allows compositionality, and ensures that the output of every algebra expression is a set of paths. Next, we provide a formal definition of this core algebra. 

\subsection{Core Algebra - Formal definition}
\label{sec:corealgebradef}
Given a path $p = (n_1,e_1,n_2,e_2,\dots,e_{k},n_{k+1})$, we define the following \emph{path operators}: 
\begin{itemize}
\item $\first(p)$: returns the identifier of the first node occurring in $p$, e.g. $\first(p) = n_1$;
\item $\last(p)$: returns the identifier of the last node occurring in $p$, e.g. $\last(p) = n_{k+1}$;
\item $\node(p,i)$: returns the identifier of the node occurring in the position $i$ of the path $p$,
e.g $\node(p,2) = n_2$; 
\item $\edge(p,j)$: returns the identifier of the edge occurring in the position $j$ of the path $p$, e.g $\edge(p,1) = e_1$;
\item $\len(p)$: returns the length (number of edges) of the path $p$, e.g. $\len(p) = k$;
\item $\lab(o)$: returns the label of an object (node or edge) $o$ occurring in $p$, e.g. $\lab(\first(p))= ``person"$.
\item $\prop(o,pr)$: returns the value of a property $pr$ of an object $o$, e.g. $\prop(\first(p),name) = ``Lisa"$.
\end{itemize}

Let $i \geq 1$ be an integer, $p$ be a path, $o \in O$ be an object, $v$ be a value, and $pr$ be a property name.
A \emph{selection condition} is defined recursively as follows. A simple selection condition is any of the expressions\footnote{Our definition of simple selection conditions can be easily extended to support inequalities ($\neq$ $<$, $>$, $\leq$ and $\geq$) and other build-in functions (e.g. $substr$ or $bound$).} $label(node(i)) = v$, $label(edge(i)) = v$, $label(first) = v$, $label(last) = v$, $node(i).pr = v$, $edge(i).pr = v$, $first.pr = v$, $last.pr = v$ and $len() = i$.
If $c_1$ and $c_2$ are selection conditions, then $(c_1  \land c_2)$, $(c_1 \lor c_2)$, and $\neg(c_1)$ are complex selection conditions.

The evaluation of a selection condition $c$ over a path $p$, denoted $ev(c,p)$, returns either $\true$ or $\false$.
A simple condition $c$ is evaluated as $\true$ in the following cases:
\begin{itemize}
\item if $c$ is $label(node(i)) = v$ and $\lab(\node(p,i))$ returns $v$;
\item if $c$ is $label(edge(i)) = v$ and $\lab(\edge(p,i))$ returns $v$;
\item if $c$ is $label(first) = v$ and $\lab(\node(p,1))$ returns $v$; 
\item if $c$ is $label(last) = v$ and $\lab(\node(\len(p) + 1))$ returns $v$;
\item if $c$ is $node(i).pr = v$ and $\prop(\node(p,i),pr)$ returns $v$;
\item if $c$ is $edge(i).pr = v$ and $\prop(\edge(p,i),pr)$ returns $v$;
\item if $c$ is $first.pr = v$ and $\prop(\node(p,1),pr)$ returns $v$;
\item if $c$ is $last.pr = v$ and $\prop(\node(p,\len(p) + 1),pr)$ returns $v$;
\item if $c$ is $len() = i$ and $\len(p)$ returns $v$.
\end{itemize} 

The evaluation of a complex selection condition is defined following the usual semantics of propositional logic.

Let $p_1$ and $p_2$ be two paths satisfying that $\last(p_1) = \first(p_2)$. The \emph{path concatenation} of $p_1$ and $p_2$, denoted $p_1 \circ p_2$, returns a new path composed by the sequence of $p_1$ followed by the tail of $p_2$ (i.e. the sequence of $p_2$ without the first node).
For example, if $p_1 = (n_1,e_1,n_2)$ and $p_2= (n_2,e_3,n_3)$ then $p_1 \circ p_2 = (n_1,e_1,n_2,e_3,n_3)$.

\begin{definition}[Core Path Algebra]
Let $\s$ and $\s'$ be sets of paths and $c$ be a selection condition. The Core Path Algebra is composed by the following operators:
\begin{itemize}
\item 
\textbf{Selection:}
$ \sigma_c(\s)= \{ p \in \s \mid ev(p,c) = \true \}$
\item 
\textbf{Join:}
$ \s \bowtie \s' = \{ p_1 \circ p_2 \mid p_1 \in \s \land p_2 \in \s' \land \last(p_1) = \first(p_2) \} $
\item 
\textbf{Union:} 
$\s \cup \s' = \{ p \mid p \in \s \lor p \in \s' \}$
\end{itemize}

\end{definition}

The intuition behind the above definition lies in the path manipulation operators. For instance, a classical operator such as the relational join whose semantics and algorithmic aspects have been widely studied in the database community would not be applicable to paths. Indeed, the conditions on the endpoints of the paths and the returned paths are not easily expressible in a relational setting. Finally, the above core algebra operators corresponds to the fundamental operators of Codd's relational algebra \cite{Codd70} under a revised 
semantics. 

\section{Recursive Path Algebra}
\label{sec:recursive}
The core algebra described in Section \ref{sec:core} allows us to express fixed-length path queries. In this section, we extend the core algebra with a recursive operator that allows retrieving paths of any length.
For example, consider that we want to obtain all the paths from the node \code{:Person} named \code{"Moe"} to the node \code{:Person} named \code{"Apu"}, through the label \code{Knows} one or more times, or through the concatenation of the labels \code{Likes} and \code{Has\_creator}, zero or more times.
This question can be answered by using the following GQL-like query: 

{ \small
\begin{verbatim}
  MATCH p = (x {name:"Moe"})-[(:Knows+)|
                 (:Likes/:Has_creator)*]->(y {name:"Apu"}) 
\end{verbatim}
}

In the above query, \code{(Knows+)} is a regular expression that allows us to obtain all the paths, of length one or more, containing edges labeled with \code{"Knows"}, and connecting the nodes \code{"Moe"} and \code{"Apu"}. 
Similarly, the regular expression \code{(Likes/Has\_creator)*} obtains the paths that combine the edge labels \code{Likes} and \code{Has\_creator}, the one following the other, an undefined number of times (including paths of length zero). This is an example of regular path query.  

Next, we define a recursive operator which allows us to evaluate regular path queries and return the entire paths.

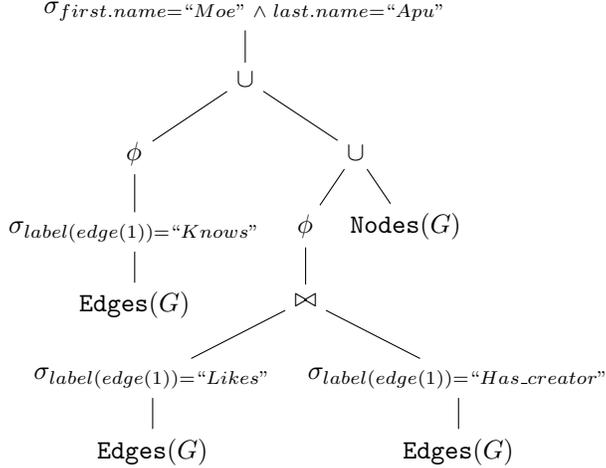
\begin{figure}[t!]
  \centering
  \begin{forest}
    [$\sigma_{first.name = ``Moe" \; \wedge \; last.name = ``Apu"}$
        [$\cup$
            [$\phi_{}$
                [$\sigma_{label(edge(1)) = ``Knows"}$
                [$\edges(G)$]
                ]
            ]
            [$\cup$
            [$\phi_{}$
                [$\bowtie$
                    [$\sigma_{label(edge(1)) = ``Likes"}$
                    [$\edges(G)$]
                    ]
                    [$\sigma_{label(edge(1)) = ``Has\_creator"}$
                    [$\edges(G)$]
                    ]
                ] 
            ]
                [$\nodes(G)$]
            ]
        ]
    ]   
\end{forest}
  \caption{Evaluation tree of a recursive path algebra query.}
  \label{fig:recursive}
\end{figure}

\begin{definition} \label{def:recop}(Recursive operator)
Given a set of paths $\s$, the recursive operator $\phi$ is defined inductively as follows:

Base case: $\phi_0(\s) = \s$.

Recursive case:

$\phi_i(\s) =  (\phi_{i-1}(\s) \bowtie \phi_0(\s)) \cup \phi_{i-1}(\s)$ while $0 < i$ and $|\phi_{i-1}| \neq |\phi_{i}|$.
\end{definition}

Note that, the recursive operator $\phi$ applies a chain of join operations over the set of paths $\s$ until the fix-point condition $|\phi_{i-1}| = |\phi_{i}|$ is reached. 
Specifically, in the base case, $\phi_0(\s)$ returns the set of paths $\s_0 = \s$. For the recursion step $i$, with $i \geq 1$, $\phi_i(\s)$ returns the set of paths $\s_{i} = \s_{i-1} \bowtie \s$, i.e. $\s_1 = \s_0 \bowtie \s$, $\s_2 = \s_1 \bowtie \s$, $\dots$, $\s_{n} = \s_{n-1} \bowtie \s$. The recursion stops when $\s_n$ is equal to $\s_0 \cup \dots \cup \s_{n-1}$, i.e. when the latest set $\s_n$ does not add new paths to the solution. Finally, $\phi(\s) = \s_0 \cup \dots \cup \s_{n-1}$. 

For example, Figure \ref{fig:recursive} shows an evaluation tree that includes the recursive operator $\phi$ twice.
In the first case (left branch), the recursive operator receives as input the paths of length one, having label \code{Knows}, which were obtained by filtering the set $\edges(G)$ (i.e. the paths of length one). Hence, the recursive operator returns the paths containing one or more edges, all of them having label \code{Knows}. This is equivalent to the result of evaluating the regular expression \code{(Knows)+}.
In the second case (right branch), the recursive operator returns paths of length 2, 4, 6, etc., such that they repeat the combination of edge labels \code{Likes} and \code{Has\_creator}. Additionally, the set of paths returned by the recursive operator is united with the set $\nodes(G)$, that contains the nodes of the graph (i.e. the paths of length zero). The result of such union is equivalent to the semantics of the Kleene star regular expression \code{(Likes/Has\_creator)*}. 

It's worth mentioning that the above definition of the recursive operator has the problem of unsolvability when the input graph contains cycles (e.g. see the inner loop formed by the label $Knows$ in the graph of Figure \ref{fig:exampleGraph}). 
In such case, the recursive operator will never halt, thereby triggering infinite results.
Fortunately, this problem can be solved by filtering the paths generated during the recursion, that is, by implementing a specific semantics for evaluating paths. 

Following the GQL specification \cite{GQL2024}, we defined five versions of the recursive operator, each one associated with a specific path semantics (i.e. walk, trail, acyclic, simple and shortest):
\begin{itemize}
\item $\phi_{Walk}$ returns all the paths without any restriction;
\item $\phi_{Trail}$ returns paths without repeated edges;
\item $\phi_{Acyclic}$ returns paths without repeated nodes;
\item $\phi_{Simple}$ returns paths without repeated nodes, with exception of the first and the last node;
\item $\phi_{Shortest}$ returns the paths with the shortest length between the first and the last node.
\end{itemize}

Note that each semantics induces a different set of solutions. For example, given the graph shown in Figure \ref{fig:exampleGraph} and the regular expression \code{Knows+}, in Table \ref{tab:paths1} we show some of the paths returned for each path semantics.
It is important to mention that, under Walk semantics, there is an infinite number of solutions due to the cycle between nodes $n_2$ and $n_3$.

There is no criteria to say which one of the above recursive operators is the best, but the corresponding semantics have been studied in theory and are implemented by practical graph query languages. Specifically, Gremlin allows arbitrary semantics, SPARQL uses acyclic path semantics, Cypher implements trail semantics, and G-CORE follows shortest path semantics.
GQL supports the above five semantics, even allowing multiple semantics in a single query. 
Next we describe how this feature is supported by our algebra. 

\begin{table}[t!]
\centering
\begin{tabular}{|c|c|c|c|c|c|c|}
\hline
\textbf{ID} & \textbf{Path} & \textbf{W} & \textbf{T} & \textbf{A} & \textbf{S} & \textbf{Sh} \\ \hline
\hline
$p_1$ & $(n_1, e_1, n_2)$ & $\checkmark$ & $\checkmark$ & $\checkmark$ & $\checkmark$ & $\checkmark$ \\ 
\hline
$p_2$ & $(n_1, e_1, n_2, e_2, n3, e_3, n_2)$ & $\checkmark$ & $\checkmark$ & & & \\
\hline
$p_3$ & $(n_1, e_1, n_2, e_2, n_3)$ & $\checkmark$ & $\checkmark$ & $\checkmark$ & $\checkmark$ & $\checkmark$ \\
\hline
$p_4$ & $(n_1, e_1, n_2, e_2, n_3, e_3, n_2, e_2, n_3)$ & $\checkmark$ & & & & \\ 
\hline
$p_5$ & $(n_1, e_1, n_2, e_4, n_4)$ & $\checkmark$ & $\checkmark$ & $\checkmark$ & $\checkmark$ & $\checkmark$ \\ 
\hline
$p_6$ & $(n_1, e_1, n_2, e_2, n_3, e_3, n_2, e_4, n_4)$ & $\checkmark$ & $\checkmark$ & & & \\ 
\hline
$p_7$ & $(n_2, e_2, n_3, e_3, n_2)$ & $\checkmark$ & $\checkmark$ & & $\checkmark$ & $\checkmark$ \\ 
\hline
$p_8$ & $(n_2, e_2, n_3, e_3, n_2, e_2, n_3, e_3, n_2)$ & $\checkmark$ & & & & \\ 
\hline
$p_9$ & $(n_2, e_2, n_3)$ & $\checkmark$ & $\checkmark$ & $\checkmark$ & $\checkmark$ & $\checkmark$ \\ \hline
$p_{10}$ & $(n_2, e_2, n_3, e_3, n_2, e_2, n_3)$ & $\checkmark$ & & & & \\
\hline
$p_{11}$ & $(n_2, e_4, n_4)$ & $\checkmark$ & $\checkmark$ & $\checkmark$ & $\checkmark$ & $\checkmark$ \\
\hline
$p_{12}$ & $(n_2, e_2, n_3, e_3, n_2, e_4, n_4)$ & $\checkmark$ & $\checkmark$ & & & \\
\hline
$p_{13}$ & $(n_3, e_3, n_2, e_4, n_4)$ & $\checkmark$ & $\checkmark$ & $\checkmark$ & $\checkmark$ & $\checkmark$ \\
\hline
$p_{14}$ & $(n_3, e_3, n_2, e_2, n_3, e_3, n_2, e_4, n_4)$ & $\checkmark$ & & & & \\
\hline
\end{tabular}
\caption{Given the graph shown in Figure \ref{fig:exampleGraph}, this table shows some (of an infinite number of) paths satisfying the regular expression \code{Knows+}, this under different semantics (W\textit{alk}, T\textit{rail}, A\textit{cyclic}, S\textit{imple} and Sh\textit{ortest}).
}
\label{tab:paths1}
\end{table}

\section{Accounting for Path Modes}
\label{sec:selres}
As described in Section \ref{sec:gql}, both standard graph query languages, i.e. GQL and SQL/PGQ, allow to express path modes (i.e. selectors and restrictors)  deciding which paths are returned and how the paths are computed, respectively. In this section, we provide a formal definition of both concepts and describe their implementation in our algebraic framework. 

Consider the following GQL query where \code{ANY SHORTEST} is the selector keyword and \code{TRAIL} the restrictor keyword:
{ 
\small
\begin{verbatim}
  MATCH ANY SHORTEST TRAIL p = (x)-[:Knows]->+(y).
\end{verbatim}
}
According to the informal descriptions for selectors (Table \ref{tab:selectors}) and restrictors (Table \ref{tab:restrictors}), the above query is evaluated as follows:
\begin{enumerate}
\item Compute the paths satisfying the regular expression \code{Knows+} by complying with the \code{TRAIL} semantics (i.e. paths without repeated edges).
\item Group the paths having the same source and target nodes.
\item Pick a single \emph{shortest} path (\code{ANY SHORTEST}) from each group, randomly (recall that there might be several trails of the same shortest length in each group).
\end{enumerate}
Hence, the restrictor allows us to define the path semantics used to compute the paths, and the selector allows us to filter out the resulting paths.

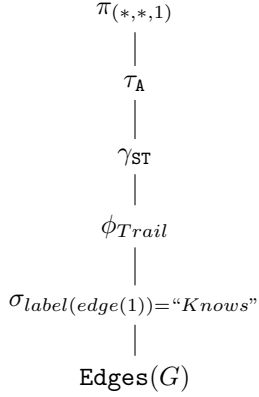
\begin{figure}[t!]
  \centering
  \begin{forest}
    [$\pi_{(*,*,1)}$
      [$\tau_{\code{A}}$
        [$\gamma_{\code{ST}}$
            [$\phi_{Trail}$
                [$\sigma_{label(edge(1)) = ``Knows"}$
                    [$\edges(G)$] 
                ]
            ]
        ]
      ]
  ]
\end{forest}
\caption{A query plan including order-by, group-by and projection.}
\label{fig:tree1}
\end{figure}

To support the query mentioned above, we introduced three algebraic operators: group-by ($\gamma$), order-by ($\tau$) and projection ($\pi$).   
To explain these operators, we will use the evaluation tree shown in Figure \ref{fig:tree1}, which emulates the GQL query described above. 
In the following, we enumerate the steps from the leaf to the root of the algebraic tree. 

\smallskip
\noindent
\textbf{Step 1.} The operator $\edges(G)$ returns the edges of the graph, that is the paths of length one.

\smallskip
\noindent
\textbf{Step 2.} The selection operator $\sigma$ filters the edges to those having label \emph{``Knows''} (i.e. $e_1$, $e_2$, $e_3$ and $e_4$).

\smallskip
\noindent
\textbf{Step 3.} The recursive operator $\phi_{Trail}$ computes the paths satisfying the regular expression \code{Knows+} by applying trail semantics. It results in the set of paths $\{ p_1, p_2, p_3, p_5, p_6, p_7, p_9, p_{11}, p_{12}, p_{13} \}$ shown in Table \ref{tab:paths1} (column $T$).

\smallskip
\noindent
\textbf{Step 4.} The group-by operator $\gamma_{\code{ST}}$ transforms the above set of paths into a solution space, a special data structured composed of partitions and groups. Specifically, a solution space is composed of one or more partitions, each partition is composed of one or more groups, and each group contains one or more paths.
A partition organize the paths based on their endpoints (i.e. the source and target nodes of a path), and a group organize the paths based on their length.
Hence, a given combination of \emph{Source}, \emph{Target} and \emph{Length} induces a solution space with a specific organization of partitions and groups as shown in Table \ref{tab:gbssdis}. 

\begin{table}[h]
\centering
\begin{tabular}{|c|c|l|}
\hline
\textbf{Group-by expression} & \textbf{Solution space organization} \\ 
\hline
$\gamma$ & 1 partition, 1 group\\
\hline
$\gamma_{\code{S}}$ & N partitions, 1 group per partition\\
\hline
$\gamma_{\code{T}}$ & N partitions, 1 group per partition\\ 
\hline
$\gamma_{\code{L}}$ & 1 partition, M groups per partition\\ 
\hline
$\gamma_{\code{ST}}$ & N partitions, 1 group per partition\\
\hline
$\gamma_{\code{SL}}$ & N partitions, M groups per partition\\ 
\hline
$\gamma_{\code{TL}}$ & N partitions, M groups per partition\\ 
\hline
$\gamma_{\code{STL}}$ & N partitions, M groups per partition\\ 
\hline
\end{tabular}
\caption{Group-by expressions and the corresponding solution space organizations.}
\label{tab:gbssdis}
\end{table}

Recalling our example, the operator $\gamma_{\code{ST}}$ (\emph{Source-Target}) implies a solution space with $N$ partitions, where each partition contains a single group with the paths having the same source and target nodes. Hence, there will be one group for each pair of people connected by a path satisfying the regular expression \code{knows+}. 
A tabular representation of this solution space is shown in Table \ref{tab:ss1}. 

\begin{table}[t!]
\centering
\begin{tabular}{|c|c|c|c|c|c|}
\hline
\textbf{Partition} $P$ & \textbf{Group} $G$ & \textbf{Path} $p$ & $\minl(P)$ & $\minl(G)$ & $\len(p)$ \\ \hline
\hline
$part_1$ & $group_{11}$ & $p_1$ & 1 & 1 & 1 \\ 
         &              & $p_2$ &   &   & 3 \\ 
\hline         
$part_2$ & $group_{21}$ & $p_3$ & 1 & 1 & 1 \\ 
\hline
$part_3$ & $group_{31}$ & $p_5$ & 2 & 2 & 2 \\ 
         &              & $p_6$ &   &   & 4 \\ 
\hline
$part_4$ & $group_{41}$ & $p_7$ & 2 & 2 & 2 \\
\hline
$part_5$ & $group_{51}$ & $p_9$ & 1 & 1 & 1 \\
\hline
$part_6$ & $group_{61}$ & $p_{11}$ & 1 & 1 & 1 \\
         &              & $p_{12}$ &   &   & 3 \\
\hline
$part_7$ & $group_{71}$ & $p_{13}$ & 2 & 2 & 2 \\
\hline
\end{tabular}
\caption{Example of solution space produced by the group-by operator $\gamma_{\code{ST}}$. The parameter \code{ST} implies many partitions, one group per partition, and many paths per group.}
\label{tab:ss1}
\end{table}

\smallskip
\noindent
\textbf{Step 5.} The order-by operator ($\tau_{\theta}$) allows us to sort the partitions, the groups and the paths composing a solution space.
The parameter $\theta$ indicates the ordering criterion, allowing the values $\code{P}$, $\code{G}$, $\code{A}$, $\code{PG}$, $\code{PA}$, $\code{GA}$ and $\code{PGA}$.
Specifically, 
$\tau_{\code{P}}$ (\emph{order-by partition}) means that the partitions are sorted by the length of their shortest path, in ascending order;
$\tau_{\code{G}}$ (\emph{order-by group}) means that the groups inside each partition are sorted by the length of their shortest path;
$\tau_{\code{A}}$ (\emph{order-by path}) means that the paths inside each group are sorted by length;
$\tau_{\code{PG}}$ (\emph{order-by partition-group}) sorts by both, partition and group; similarly for the remaining cases.

Following with our example, the operation $\tau_{\code{A}}$ sorts the paths inside each group of the solution space produced by the group-by operator.
It can be observed in Table \ref{tab:ss1}, where the column $\minl(P)$ indicates the length of the shortest path in a partition $P$, $\minl(G)$ indicates the length of the shortest path in a group $G$, and $\len(p)$ indicates the length of a path $p$. Note that these columns can be used to sort partitions, groups and paths. 

\smallskip
\noindent
\textbf{Step 6}. The projection operator ($\pi$) allows us to transform a solution space into a set of paths.
To do this, the projection operator receives as parameter a tuple of the form $(\#_{P}, \#_{G}, \#_{A})$ where each $\#$ can be either the symbol $*$ or a positive integer. Hence, $\#_{P}$ indicates the number of partitions to be returned, $\#_{G}$ indicates the number of groups (per partition) to be returned, and $\#_{A}$ indicates the number of paths (per group) to be returned. 

Coming back to our example, the expression $\pi_{(*,*,1)}$ returns one path per group (the first one in the group; that is, the shortest one since in the previous step we sorted the paths by length), for every group inside each partition. Hence, the final query output will be the set of paths $\{ p_1, p_3, p_5, p_7, p_9, p_{11}, p_{13} \}$. 

In the remainder of this section, we study the formal semantics for the algebraic operators group-by, order-by and project, all of which use the notion of solution space. 

\begin{definition}
A \emph{Solution Space} is a tuple  
$SS = (\s, \mathbb{G}, \mathbb{P}, \ptog, \gtop, \ord)$ 
where: 
$\s$ is a set of paths; 
$\mathbb{P}$ is a set of partitions; 
$\mathbb{G}$ is a set of groups; 
$\ptog: \s \to \mathbb{G}$ is a total function that assigns each path to a group; 
$\gtop: \mathbb{G} \to \mathbb{P}$ is a total function that assigns each group to a partition;
and 
$\ord : (\s \cup \mathbb{G} \cup \mathbb{P}) \to \mathbb{Z}^+$ is a total function used to assign a positive integer to paths, groups and partitions.
\end{definition}

The concept of a solution space involves a data structure that organizes a set of paths into groups (function $\ptog$), which are further organized into partitions (function $\gtop$). Additionally, the function $\ord$ is used to sort the elements within the solution space. 
For example, assume that $x$, $y$ and $z$ are paths within a group $g$.
If $\ord(x) = 1$, $\ord(y) = 2$ and $\ord(z) = 3$, then we establish a virtual order of the paths inside $g$.
On the hard, if the three paths have the same value for $\ord$, then there is no order among them.
The same approach is used to sort the groups within a partition and to sort the partitions within the solution space.     

\subsection{Group by}
Given a set of paths $\s$, the group-by operator is represented as $\gamma_{\psi}(\s)$ where $\psi \in \{\emptyset, \code{S}, \code{T}, \code{L}, \code{ST}, \code{SL}, \code{TL}, \code{STL} \}$ (\code{S} = Source, \code{T} = Target, \code{L} = Length, \code{ST} = Source-Target, \code{SL} = Source-Length, \code{TL} = Target-Length, \code{STL} = Source-Target-Length).
The evaluation of $\gamma_{\psi}(\s)$ returns an solution space
$SS = (\s, \mathbb{G}, \mathbb{P}, \ptog, \gtop, \ord)$
defined as follows:
\begin{itemize}
\item 
If $\psi$ is $\emptyset$ then $\mathbb{P}=\{P_1\}$, $\mathbb{G}=\{G_1\}$, $\gtop(G_1)=P_1$, and $\forall p \in \s$ it applies that $\alpha(p) = G_1$.
\item 
If $\psi$ is \code{S} then $\forall s \in \{ \first(p) \mid p \in \s \}$ there will be a partition $P_s = \{ G_s \}$ where $G_s = \{ p' \in \s \mid \first(p') = s \}$.
\item 
If $\psi$ is \code{T} then $\forall t \in \{ \last(p) \mid p \in \s \}$ there will be a partition $P_t = \{ G_t \}$ where $G_t = \{ p' \in \s \mid \last(p') = t \}$.
\item 
If $\psi$ is \code{L} then $\mathbb{P}=\{P_1\}$, $\forall l \in \{ \len(p) \mid p \in \s \}$ there will be a group $G_l = \{ p' \in \s \mid \len(p') = l \}$ with $\gtop(G_l) = P_1$.
\item 
If $\psi$ is \code{ST} then $\forall (s,t) \in \{ (\first(p),\last(p)) \mid p \in \s \}$ there will be a partition $P_{st} = \{ G_{st} \}$ where $G_{st} = \{ p' \in S \mid \first(p') = s \land \last(p') = t \}$.
\item 
If $\psi$ is \code{SL} then $\forall s \in \{ \first(p) \mid p \in \s \}$ there will be a partition $P_s$, and $\forall l \in \{ \len(p) \mid p \in \s \land \first(p) = s \}$ there will be a graph $G_{sl} = \{ p' \in \s \mid \first(p') = s \land \len(p') = l \}$ with $\gtop(G_{sl}) = P_s$.
\item 
If $\psi$ is \code{TL} then $\forall t \in \{ \last(p) \mid p \in \s \}$ there will be a partition $P_t$, and $\forall l \in \{ \len(p) \mid p \in \s \land \last(p) = t \}$ there will be a graph $G_{tl} = \{ p' \in \s \mid \last(p') = t \land \len(p') = l \}$ with $\gtop(G_{tl}) = P_t$.
\item 
If $\psi$ is \code{STL} then $\forall (s,t) \in \{ (\first(p),\last(p)) \mid p \in \s \}$ there will be a partition $P_{st}$, and $\forall l \in \{ \len(p) \mid p \in S \land \first(p) = s \land \last(p) = t \}$ there will be a graph $G_{stl} = \{ p' \in \s \mid \first(p') = s \land \last(p') = t \land \len(p') = l \}$ with $\gtop(G_{stl}) = P_{st}$.

\end{itemize}

Additionally, it applies that: $\ord(p) = 1$ for every path $p \in \s$, $\ord(G) = 1$ for every group $G \in \mathbb{G}$, and $\ord(P) = 1$ for every partition $P \in \mathbb{P}$. Recall that function $\ord$ can be used to introduce a virtual order of the elements (paths, groups and partitions) within a solution space. 
In this case, there is no such order as all the elements has the same value for $\ord$.

\subsection{Order by}
Let $SS = (\s, \mathbb{G}, \mathbb{P}, \ptog, \gtop, \ord)$ be a solution space. Given a group $G \in \mathbb{G}$, we will use the function $\minl(G)$ to obtain the length of the shortest path in $G$. Similarly, for a given partition $P \in \mathbb{P}$, the function $\minl(P)$ returns the minimum length among all the groups in $P$. 

Given a solution space $SS$, the order-by operator is represented as $\tau_{\theta}(SS)$  where $\theta \in \{ \code{P}, \code{G}, \code{A}, \code{PG}, \code{PA}, \code{GA}, \code{PGA} \}$ (\code{P} = Partition, \code{G} = Group, \code{A} = Path, \code{PG} = Partition-Group, \code{PA} = Partition-Path, \code{GA} = Group-Path, \code{PGA} = Partition-Group-Path).
The evaluation semantics of $\tau_{\theta}(SS)$ is presented in Table \ref{tab:obsem}.
Given a solution space $SS = (\s, \mathbb{G}, \mathbb{P}, \ptog, \gtop, \ord)$, the operator $\tau_{\theta}(SS)$ redefines the function $\ord$ to a new function $\ord'$ depending on the parameter $\theta$.
For instance, if $\theta$ is \code{P} then, for each partition $P$ it applies that $\ord'(P) = \minl(P)$, for each group $G$ in $P$ it applies that $\ord'(G) = \ord(G)$, and for each path $p$ in $G$ it applies that $\ord'(p) = \ord(p)$.

\begin{table}[t!]
\centering
\begin{tabular}{|l|l|l|l|}
\hline
$\theta$ & $\forall P \in \mathbb{P}$ & $\forall G \in \mathbb{G}$ & $\forall p \in \s$ \\
\hline
\code{P} & $\ord'(P) = \minl(P)$ & $\ord'(G) = \ord(G)$ & $\ord'(p) = \ord(p)$ \\
\hline
\code{G} & $\ord'(P) = \ord(P)$ & $\ord'(G) = \minl(G)$ & $\ord'(p) = \ord(p)$ \\
\hline
\code{A} & $\ord'(P) = \ord(P)$ & $\ord'(G) = \ord(G)$ & $\ord'(p) = \len(p)$ \\
\hline
\code{PG} & $\ord'(P) = \minl(P)$ & $\ord'(G) = \minl(G)$ & $\ord'(p) = \ord(p)$ \\
\hline
\code{PA} & $\ord'(P) = \minl(P)$ & $\ord'(G) = \ord(G)$ & $\ord'(p) = \len(p)$ \\
\hline
\code{GA} & $\ord'(P) = \ord(P)$ & $\ord'(G) = \minl(G)$ & $\ord'(p) = \len(p)$ \\
\hline
\code{PGA} & $\ord'(P) = \minl(P)$ & $\ord'(G) = \minl(G)$ & $\ord'(p) = \len(p)$ \\
\hline
\end{tabular}
\caption{Semantics of the order-by operator $\tau_{\theta}$. Given a solution space 
$SS = (\s, \mathbb{G}, \mathbb{P}, \ptog, \gtop, \ord)$, the evaluation of $\tau_{\theta}(SS)$ returns a solution space
$SS' = (\s, \mathbb{G}, \mathbb{P}, \ptog, \gtop, \ord')$ where function $\ord'$ is the only change. For each value of the parameter $\theta$, this table shows the assignments for function $\ord'$.
}
\label{tab:obsem}
\end{table}

Note that the order-by operator uses the function $\ord'$ to introduce a virtual ordering of paths, groups and partitions. The order of a path, inside a group, is given by its length. The order of a group, inside a partition, is given by the length of its shortest path. The order of a partition, inside a solution space, is given by the length of the shortest path contained in its groups.

\subsection{Projection}
Given an solution space 
$SS = (\s, \mathbb{G}, \mathbb{P}, \ptog, \gtop, \ord)$,
the projection operator is represented as $\pi_{(\#_{P},\#_{G},\#_{A})}(SS)$ where each $\#$ is either the symbol $*$ or a positive integer. Here, $\#_{P}$ indicates the number of partitions to be projected, $\#_{G}$ indicates the number of groups (per partition) to be projected, and $\#_{A}$ indicates the number of paths (per group) to be projected.

The evaluation of $\pi_{(\#_{P},\#_{G},\#_{A})}(SS)$ returns a set of paths according to Algorithm \ref{alg:proj}. 
The final set of paths will be stored in variable $\s_{out}$.
First, the algorithm transforms the set of partitions $\mathbb{P}$ into the sequence $SeqP$ where the partitions  are sorted in ascending order according to their length defined by function $\ord$ (line 2). 
The number of partitions to be processed is defined in  variable $maxP$ according to the parameter $\#_{P}$ (lines 3 and 4).
For each partition $P$ (until processing $maxP$ partitions), the algorithm creates a sequence $SeqG$ containing the groups of $P$ (lines 6 to 8). The groups in $SeqG$ are sorted in ascending order according to function $\ord$.
The variable $maxG$ indicates the number of groups to be processed according to the parameter $\#_{G}$ (lines 9 and 10).
For each group $G$ (until processing $maxG$ groups), the algorithm gets the paths of $G$ and creates the sequence $SeqS$, where the paths are sorted based on their length (lines 12 to 14).
The variable $maxS$ indicates the number of paths to be processed according to the parameter $\#_{A}$ (lines 15 and 16).
For each path $p$ (until processing $maxS$ paths), the algorithm adds $p$ to the final set of paths $\s_{out}$ (lines 17 to 20).

Note that the procedure shown in Algorithm \ref{alg:proj} can be improved in several ways: the $Sort$ function can implement an special sorting algorithm; indexes can be used to facilitate the retrieval of paths inside a group, as well as the groups inside a partition; the calls to ordering functions (lines 2, 8 and 14) are unnecessary when the input solution space comes from the group-by operator (i.e. the query does not include the order-by operator).
Moreover, Algorithm \ref{alg:proj} can be easily extended to support the projection of partitions, groups and paths in descending order.

\begin{algorithm}
\caption{Projection function $\pi_{(\#_{P},\#_{G},\#_{A})}(SS)$}\label{alg:proj}
\SetKwInOut{Input}{input}\SetKwInOut{Output}{output}
\Input{A solution space $SS = (\s, \mathbb{G}, \mathbb{P}, \ptog, \gtop, \ord)$ and the projection parameters $\#_{P}$, $\#_{G}$ and $\#_{A}$.}
\Output{A set of paths $\s_{out}$.}
$\s_{out} \gets \emptyset$\; 
$SeqP \gets Sort(\mathbb{P})$; //Sort $\mathbb{P}$ based on function $\ord$\\
\lIf{$\#_{P}$ is \code{"*"} $\lor$ $\#_{P} > length(SeqP)$}{ $maxP \gets length(SeqP)$}
\lElse{ $maxP \gets \#_{P}$}
\For{$i\leftarrow 1$ \KwTo $maxP$}{
  $P \gets SeqP[i]$\;
  $SetG \gets \{ G \in \mathbb{G} \mid \gtop(G) = P \}$; //Get the groups of $P$\\
  $SeqG \gets Sort(SetG)$; //Sort the groups of $P$ based on $\ord$\\
  \lIf{$\#_{G}$ is \code{"*"} $\lor$ $\#_{G} > length(SeqG)$}{ $maxG \gets length(SeqG)$}
  \lElse{ $maxG \gets \#_{G}$}
  \For{$j\leftarrow 1$ \KwTo $maxG$}{
    $G \gets SeqG[j]$\;
    $SetS \gets \{ p \in S \mid \ptog(p) = G \}$; //Get the paths of $G$\\
    $SeqS \gets Sort(SetS)$; //Sort the paths of $G$ based on $\ord$\\
    \lIf{$\#_{A}$ is \code{"*"} $\lor$ $\#_{A} > length(SeqS)$}{ $maxS \gets length(SeqS)$}
    \lElse{ $maxS \gets \#_{A}$}
    \For{$k\leftarrow 1$ \KwTo $maxS$}{
      $p \gets SeqS[k]$\;
      $Add(p,\s_{out})$ //Add the path $p$ to the set $S_{out}$
    }
  }
}
\end{algorithm}

\section{Comparison with GQL}
Previously, we have described the seven types of selectors and the four types of restrictors supported by GQL. 
According to the GQL specification, it is possible to combine every selector with every restrictor, resulting in 28 combinations. 
An important fact is that every selector-restrictor combination can be translated into a path algebra expression, whose evaluation satisfies the informal semantics defined in Table \ref{tab:selectors} and Table \ref{tab:restrictors}. 

In Table \ref{tab:gql-vs-alg}, we present the GQL expressions produced by combining the seven selectors with the restrictor \code{WALK}, and show the corresponding path algebra expressions.
Note that:
\code{RE} denotes the regular expression obtained from the path pattern expression $ppe$;
every selector (e.g. \code{ALL}) is translated to an expression containing the group-by ($\gamma$), order-by ($\tau$) and projection ($\pi$) operators; and
the restrictor \code{WALK} is translated to the recursive operator $\phi_{Walk}$.
The path algebra expressions shown in Table \ref{tab:gql-vs-alg}, can be replicated to translate the rest of restrictors by just replacing the term \code{WALK} with \code{TRAIL}, \code{ACYCLIC} or \code{SIMPLE}.

\begin{table}[t!]
\centering
\begin{tabular}{|l|l|}
\hline
\textbf{GQL expression} & \textbf{Path algebra expression} \\ 
\hline
\code{ALL} \code{WALK} $ppe$ & 
$\pi_{(*,*,*)}(\gamma(\phi_{Walk}(\code{RE})))$  \\ 
\hline
\code{ANY SHORTEST WALK} $ppe$ & 
$\pi_{(*,*,1)}(\tau_{\code{A}}(\gamma_{\code{ST}}(\phi_{Walk}(\code{RE}))))$  \\ 
\hline
\code{ALL SHORTEST WALK} $ppe$ & 
$\pi_{(*,1,*)}(\tau_{\code{G}}(\gamma_{\code{STL}}(\phi_{Walk}(\code{RE}))))$  \\ 
\hline
\code{ANY WALK} $ppe$ & 
$\pi_{(*,*,1)}(\gamma_{\code{ST}}(\phi_{Walk}(\code{RE})))$  \\ 
\hline
\code{ANY} $k$ \code{WALK} $ppe$ & 
$\pi_{(*,*,k)}(\gamma_{\code{ST}}(\phi_{Walk}(\code{RE})))$  \\ 
\hline
\code{SHORTEST} $k$ \code{WALK} $ppe$ & 
$\pi_{(*,*,k)}(\tau_{\code{A}}(\gamma_{\code{ST}}(\phi_{Walk}(\code{RE}))))$  \\ 
\hline
\code{SHORTEST} $k$ \code{GROUP} \code{WALK} $ppe$ & 
$\pi_{(*,k,*)}(\tau_{\code{G}}(\gamma_{\code{STL}}(\phi_{Walk}(\code{RE}))))$  \\
\hline
\end{tabular}
\caption{For each GQL selector-restrictor expression, this table shows the corresponding path algebra expression.}
\label{tab:gql-vs-alg}
\end{table}

For instance, the GQL expression
{ \small 
\begin{verbatim}
  MATCH ALL SHORTEST ACYCLIC p = (x)-[:Knows]->+(y)
\end{verbatim}
}
\noindent
can be translated to the path algebra expression
\begin{center}
$\pi_{(*,1,*)}(\tau_{\code{G}}(\gamma_{\code{STL}}(\phi_{Acyclic}(\sigma_{label(edge(1)) = ``Knows"}(\edges(G))))))$
\end{center}
where:
$\phi_{Acyclic}$ returns a set of paths where each path $p$ satisfies the regular expression \code{Knows+} and $p$ does not have any repeated nodes;
$\gamma_{\code{STL}}$ transforms the set of paths into a solution space where, for each source-target combination there is a partition $P_i$, and each partition $P_i$ contains a group $G_{ij}$ containing the paths with the same length;
$\tau_{\code{G}}$ sorts the groups inside each partition, in increasing order, according to their length (in this case, the length of each group is given by the length of the shortest path inside it); and
$\pi_{(*,1,*)}$ returns the paths contained in the first group of each partition (i.e. all the shortest paths for each source-target combination).

On the other side, there exist path algebra expressions that are not supported by GQL. It is given by the 8 types of group-by, 7 types of order-by, the 7 types of projection, and the 5 types of recursion, which results in 1960 combinations, surpassing the 28 combinations defined by GQL.
For instance, the following algebra expression is not supported by GQL:
\begin{center}
$\pi_{(*,*,1)}(\tau_{\code{G}}(\gamma_{\code{L}}(\phi_{Trail}(\sigma_{label(edge(1)) = ``Knows"}(\edges(G))))))$.
\end{center}

Note that, 
$\phi_{Trail}$ computes the trails satisfying \code{Knows+};
$\gamma_{\code{L}}$ creates a solution space with a single partition and many groups where each group contains the paths with the same length; 
$\tau_{\code{G}}$ sorts the groups in ascending order according to their length (i.e. the length of a group is equal to the length of the paths contained in the group); and 
$\pi_{(*,*,1)}$ returns a single path from each group. 
Therefore, the above algebra expression returns a sample trail of each possible length.

Another issue is the practical use of some path algebra expressions. For instance, the expression 
\begin{center}
$\pi_{(*,*,1)}(\tau_{\code{PG}}(\gamma(\phi_{Walk}(\sigma_{label(edge(1)) = ``Knows"}(\edges(G))))))$
\end{center}
is somehow redundant and unnecessarily complex as we just like to return a single path (indeterministic), without imposing any condition. Note that the order-by operator $\tau_{\code{PG}}$ is unnecessary as the operator $\gamma$ returns a solution space with a single partition and a single group. Hence, the above expression can be replaced by 
\begin{center}
$\pi_{(*,*,1)}(\gamma(\phi_{Trail}(\sigma_{label(edge(1)) = ``Knows"}(\edges(G)))))$.
\end{center}

For the sake of space, we are not presenting a complete analysis of the equivalences among the path expressions supported by our path algebra and GQL. However, the above examples give an intuition of the expressive power of our algebra, and show that it supports more types of queries that the ones supported by GQL. In terms of completeness, 
our algebra allows for all possible combinations of its operators, leaving the responsibility of using them correctly to the user. This approach is also followed in practical query languages like SQL and Cypher.


\section{Connecting the path algebra with GQL}
Our main motivation to define a path algebra is to allow formulating logical plans for the execution of path queries in a fashion similar to the usage of relational algebra for SQL logical plans. Most notably, having a path algebra at our disposal allows for a quick formulation of logical plans in any engine wishing to support queries that involve complex conditions over paths. 
To aid with this process, we extended the GQL syntax to  support all the features provided by the algebra, and implemented a query parser for it.

\subsection{GQL Extension}
According to the GQL specification, the structure of a path query is given by the following grammar:

{ \small
\begin{verbatim}
<pathQuery> ::= MATCH <selector> <restrictor> 
                <pathPattern>
<selector> ::= ALL | ANY SHORTEST | ALL SHORTEST | 
               ANY | ANY <number> | 
               SHORTEST <int> | SHORTEST <int> GROUP 
<restrictor> ::= WALK | TRAIL | SIMPLE | ACYCLIC
<pathPattern> ::= <var> = <pathExp> WHERE <condition> 
\end{verbatim}
}
\noindent
where \code{<pathExp>} is an expression of the form 
\code{(var)-[RegExp]->(var)}, and \code{<condition>} is a selection condition.

In order to support the operators of our path algebra, we propose to modify the above structure as follows:   

{ \small
\begin{verbatim}
<pathQuery> ::= MATCH <projection> <restrictor_ext> 
                <pathPattern> <groupby>? <orderby>?
<projection> ::= <partProj> <groupProj> <pathProj>
<partProj> ::= ( ALL | <number> ) PARTITIONS
<groupProj> ::= ( ALL | <number> ) GROUPS
<pathProj> ::= ( ALL | <number> ) PATHS
<restrictor_ext> ::= WALK | TRAIL | SIMPLE | ACYCLIC | SHORTEST                      
<groupby> ::= (SOURCE)? (TARGET)? (LENGTH)?
<orderby> ::= (PARTITION)? (GROUP)? (PATH)?
\end{verbatim}
}

Hence, if we want to compute all the trails for each pair of nodes in the graph, and return a single path for each target node, we can use the query

{ \small
\begin{verbatim}
  MATCH ALL PARTITIONS ALL GROUPS 1 PATHS 
  TRAIL p = (?x)-[(:Knows)*]->(?y) 
  GROUP BY TARGET ORDER BY PATH
\end{verbatim}
}

\noindent
which corresponds to the path algebra expression
\begin{center}
$\pi_{(*,*,1)}(\tau_{\code{A}}(\gamma_{\code{T}}(\phi_{Trail}(\sigma_{label(edge(1)) = ``Knows"}( 
 \edges(G))))))$.
\end{center}

\subsection{Query Parser}
We have developed\footnote{https://github.com/pathalgebra/AlgebraParser} an open-source parser that transforms declarative queries into logical query plans.
Our parser is a Java application with two main components: the query parser and the logical plan generator.
The query parser reads a path query expression and returns a parse tree. This is achieved by using the \emph{ANTLR}\footnote{ANTLR (ANother Tool for Language Recognition), https://www.antlr.org} library. 
The logical plan generator traverses the parse tree, extracting all the algebraic operations, and generates a query tree.

The query parser has a command-line interface where users can write a path query expression and get the corresponding parse tree. For example, if we input the sample query shown above in this section, the parser generates the following query plan:
\begin{Verbatim}[numbers=left]
Projection (ALL PARTITIONS ALL GROUPS 1 PATHS)
OrderBy (Path)
Group (Target)
Restrictor (TRAIL)
-> Recursive Join (restrictor: TRAIL)
 -> Select: (label(edge(1)) = "Knows" , EDGES(G))
\end{Verbatim}

The output of the parser displays the query plan as a textual tree. The initial lines (1 to 4) display the selected parameters for the projection, order by, group by, and restrictor statements. Subsequently, lines 5 and 6 present the query, with indentation indicating the depth of each instruction and its corresponding branch.

Notice that our logical plans pave the way for building implementations of GQL and SQL/PGQ standards, or for general engines wishing to support path queries. Namely, to build a reference implementation, one only needs to specify an algorithm for each operator of the algebra, as these suffice to define any path query in the two standards (and some additional ones as well). Notice that algorithms for specific algebra operations we support are independent research topics of their own~\cite{YakovetsGG16,Farias2023ARXIV,WoldeSB23}, so we find building such a reference implementation to be outside of the scope of the current paper.

\subsection{Query optimization} 
A well-know advantage of having a query algebra is that it facilitates query optimization. Particularly, the query engine can perform logical optimizations (e.g., predicate pushdown, column pruning) and physical optimizations (e.g., better join strategies).

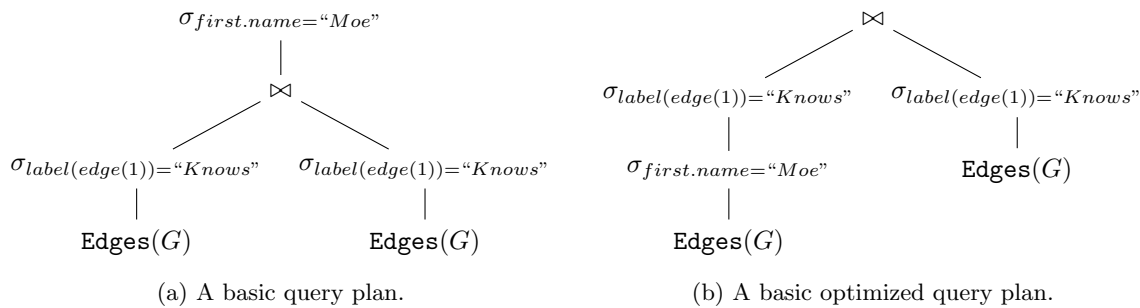
\begin{figure}[t!]
    \begin{subfigure}{.5\textwidth}
          \centering
          \begin{forest}
            [$\sigma_{first.name = ``Moe"}$
                [$\bowtie$
                    [$\sigma_{label(edge(1)) = ``Knows"}$
                        [$\edges(G)$]
                    ]
                    [$\sigma_{label(edge(1)) = ``Knows"}$
                        [$\edges(G)$]
                    ]
                ] 
            ]
        \end{forest}
        \caption{A basic query plan.}
        \label{fig:qtreeBALgebraOptA}
    \end{subfigure}
        \begin{subfigure}{.5\textwidth}
        \centering
            \begin{forest}
                [$\bowtie$
                    [$\sigma_{label(edge(1)) = ``Knows"}$
                        [$\sigma_{first.name = ``Moe"}$
                            [$\edges(G)$]
                        ]
                    ]
                    [$\sigma_{label(edge(1)) = ``Knows"}$
                        [$\edges(G)$]
                    ]
                ] 
            \end{forest}
            \caption{A basic optimized query plan.}
            \label{fig:qtreeBALgebraOptB}
        \end{subfigure}%
    \caption{A basic optimized query plan for a query with basic algebra expression.}
  \label{fig:qtreeBALgebraOpt}
\end{figure}

The classical example of logical optimization is pushing filters~\cite{garcia2008database}. For example, the query plan show in Figure \ref{fig:qtreeBALgebraOptA} can be optimized by pushing down the selection $\sigma_{first.name = ``Moe"}$.
The optimized query plan is shown in Figure \ref{fig:qtreeBALgebraOptB}.
Note that this change allows us to reduce the number of intermediate results (paths) in advance, and consequently, reduce the number of join comparisons. 

The introduction of selectors and restrictors have opened the door to new types of rewritings. 
For instance, the expression
\begin{center}
$\pi_{(1,1,*)}(\tau_{\code{G}}(\gamma_{\code{L}}(\phi_{Walk}(\sigma_{label(edge(1)) = ``Knows"}( 
 \edges(G)))))))$
\end{center}
can be used to obtain the shortest paths satisfying the regular expression $\code{Knows+}$. However, this expression just works well when the target graph does not contain cycles for the edges labeled with $\code{Knows}$. The equivalence also applies if the implementation of $\phi_{Walk}$ limits the recursion to a given length of paths, long enough to return all the shortest paths. Otherwise, the equivalence does not apply. 
Instead of the above, the optimization engine can produce the algebra expression 
\begin{center}
$\pi_{(1,1,*)}(\gamma(\phi_{Shortest}(\sigma_{label(edge(1)) = ``Knows"}( 
 \edges(G))))))$
\end{center}
where the recursive operator and the group-by operator have been changed, and the order-by operator has been removed. The change of $\phi_{Walk}$ by $\phi_{Shortest}$ is very important because now the query returns a finite number of solutions, i.e. it always terminates. 

Overall, such manipulations are a standard part of any cost-based query execution plan in SQL databases, and have a high potential to be used over graphs. Indeed, some engines already considered such optimizations for restricted versions of path algebra, or for simply detecting, but not returning paths~\cite{mdb,Yakovets2016}, and there is a rich body of literature on query rewriting for path queries~\cite{CalvaneseGLV03}.

\section{Related Work}
Since the pioneering work of Edgar Codd \cite{Codd70}, relational algebra has been 
a hallmark of data management research focusing on relational query processing and optimization. 
Despite the fact that graph databases have established themselves as a data-driven technology with high demand in industry, a counterpart of relational algebra is absent in graph database research. 
In this paper, we fill this gap and propose a path-based algebraic framework, that seamlessly work for recursive graph queries. 

In the following, we discuss related work in this area, focusing primarily on extensions of relational algebra to support path queries, algorithmic approaches for computing path queries, and the methods used by current database systems.

\subsection{Extensions of relational algebra}
First, we review extensions of relational algebra that support the evaluation of path queries.
$\mu$-RA \cite{JachietGGL20} shows the usage of $\mu$ calculus to enhance relational algebra with a recursive operator $\mu$ but they do not consider path-oriented semantics. 

GPC (Graph Pattern Calculus) \cite{FrancisGGLMMMPR23} is a declarative non-procedural calculus for property graphs similar in spirit to TRC (tuple-relational calculus) for relational data. Graph patterns in GPC generalize conjunctive two-way regular path
queries \cite{Bonifati2018} to property graphs. Despite supporting restrictors among the set of simple, trail (used by
default if none is given) and shortest, GPC returns a set of bindings along with a set of witnessing paths for those bindings. However, it cannot manipulate paths as we propose in our algebraic framework. 

There exist early attempts to define non-recursive algebras for property graph queries \cite{2018Bonifati,PacaciBO22,10.1007/978-3-031-15740-0_6}.
All these attempts closely resemble relational algebra as they redefine selection, node-based join, edge-based join and union in a similar fashion and add a path navigation operator, which is meant to encode linear recursion in RPQ. 
They disregard the definition of a path-oriented algebraic framework equipped with a projection operator and a solution space allowing for full expressiveness, covering recent graph standard query languages and beyond.  
Moreover, they disregard several features of standard graph query languages, such as restrictors and selectors, and they do not map to concrete and practical standard query languages. 

There exist other pieces of work where the notion of path algebras is used, such as case studies of graph problems involving paths \cite{Gondran1975} (connectivity, shortest path, path enumeration, among others). Searching for paths between a pair of nodes using operators for paths like join and product has also been studied \cite{Manger2004}, along with recursive operators for paths with their evaluation based on automata \cite{90388}. However, these studies are prior to the appearance of property graph-related languages and are tailored to plain labeled graphs and non-recursive queries. 
    
It is worth mentioning that there is a substantial body of work on processing SPARQL property path using an algebraic approach. Some of these~\cite{YakovetsGG16,LibkinRSV18,ReutterSV21} include defining an intermediate algebra 
supporting recursion. However, apart from the inherent differences between RDF and property graphs, they are not considering how paths should be returned, since this is not supported in SPARQL, and they do not allow any path mode besides ANY WALK.

\subsection{Algorithmic approaches for computing path queries}
The literature presents several algorithmic approaches for computing path queries in graph databases.
The most basic approach is to extend a graph traversal algorithm, such as depth-first search or breadth-first search, by incorporating regular expression matching during the traversal. 
This approach can be improved by using several techniques, including parallelism \cite{Miura2019}, approximation \cite{Wadhwa2019}, distributed processing \cite{Guo2021}, and compact structures \cite{Arroyuelo2023}.

Automata-based approaches traverse the graph while tracking the states of an automaton constructed from the regular expression, as a set of transitions maps directly to paths in the graph \cite{mendelzon1995}.
Index-based approaches precompute and index paths based on their edge labels or combinations of labels, which reduces the search space and can accelerate path computation \cite{fletcher2016,Kuijpers2021}.
Matrix-based methods represent the graph as an adjacency matrix, enabling the use of matrix multiplication to find paths between nodes \cite{arroyuelo2023b}.

Works such as~\cite{Farias2023ARXIV,MartensNPRVV23} describe specialized algorithms for executing a single path query, or compressing the paths in the result set, but they do not discuss on how these solutions can be incorporated into a larger query pipeline, nor how to specify an algebra to manipulate the output paths, unlike our approach.

These algorithmic approaches vary in efficiency, scalability, and complexity, and are often selected based on the specific characteristics of the graph and the query workload. For instance, automata-based approaches are particularly effective for graphs with clear label patterns, while index-based and matrix-based methods tend to be more efficient for large-scale graphs

An important disadvantage of using an algorithm is that we are not able to apply query optimization techniques. For doing so, we need a query algebra like the ones described in \cite{Yakovets2016} and \cite{Jachiet2020}. A common issue of these algebras is that they do not return the entire paths, just the source and target nodes for each path.

\subsection{Path query evaluation in current database systems}
We conducted a brief review of current database systems to understand the methods used for evaluating path queries.
Our revision included the following systems: Amazon Neptune, ArangoDB, DuckDB, GraphScope, MemGraph, MillenniumDB, NebulaGraph, Neo4j, OraclePGX, OrientDB, RedisGraph, TigerGraph, and Kuzu.
Next, we present key representative findings from our review. 

DuckPGQ~\cite{WoldeSB23} is one of the few systems beyond Oracle PGX implementing SQL/PGQ path queries, the first version of one of the standard query languages for property graphs. It relies on a relational backend and deals with recursion either by unfolding it to several joins depending on the length of the paths or by resorting to multi-source BFS as an external module. However, the supported fragment of SQL/PGQ path queries is rather limited and it includes only the ANY SHORTEST WALK path mode. DuckPGQ implements a version of relational algebra enhanced with UDFs (user-defined functions) to support path queries evaluation. 

MillenniumDB~\cite{mdb} supports all path modes of GQL with full RPQs, but only on a level of a single path. Combining sets of paths according to a formal algebraic framework goes beyond the current scope of MilleniumDB.

Neo4j~\cite{Webber12} offers full support for finding trails  and walks, but does not support arbitrary regular expressions to define all regular path queries. On the other hand, since the Cypher query language supports post-processing of the returned paths (viewed as lists)~\cite{Francis2018}, non-recursive algebraic operations can be simulated in the system, but are not natively implemented.

Oracle/PGX~\cite{RestHKMC16} is a graph extension of the Oracle relational backend. It relies on Compressed Sparse Row storage and is capable of evaluating conjunctive RPQs under Walk, Trail, Simple and Acyclic semantics.  

Kùzu~\cite{kuzudbRecursiveRelationship} fully supports the walk semantics, but not for all regular expressions, and it limits the path length to 30. Trails and Acyclic walks are also supported. Similarly to Cypher, some post processing can be done on retrieved paths (which are stored as lists) in order to simulate non-recursive algebraic operations.

In conclusion, none of the current database systems incorporate a path-based algebraic query evaluation as presented in this work. Supporting such an algebra would facilitate the implementation of logical plans and the application of query optimization techniques in current and future graph database systems.     

\section{Conclusions and future work}
One of the key characteristics of graph queries is the ability to return paths instead of relations. Future versions of graph query languages, defined as part of the ISO/IEC standardization activities, will be able to manipulate paths and to ensure composability of queries. 
Our work is a considerable milestone towards this direction by providing a graph algebra that lays the foundations of 
composable graph queries. Such an algebra contains fundamental operators as in Codd's relational algebra but goes significantly beyond them by defining recursive operators and coverage of path modes (selectors and restrictors) defined in GQL and SQL/PGQ. The latter are defined by specifying grouping, order by and projection operators on path variables. Our work is accompanied by a parser implementing the algebraic operators. 

Our work paves the way to further research on graph query algebraic optimizations leading to efficient implementations of 
graph query engines. While current graph databases including open-source and commercial ones are adding features from the standards (GQL and SQL/PGQ), they also need to provide internals for logical and physical plans inspired by a formal algebraic framework, as the one proposed in our paper. 

\paragraph{Acknowledgments.}
This work was supported by ANID FONDECYT Chile through grant 1221727.
R. García was supported by  CONICYT-PFCHA/Doctorado Nacional/2019-21192157. A. Bonifati was supported by ANR VeriGraph (nr. ANR-21-CE48-0015). D. Vrgo\v{c} was supported by FONDECYT Regular grant number 1240346.

\bibliographystyle{acm}
\bibliography{paper}

\end{document}